\documentstyle[12pt]{article}
\normalbaselines 
\newcommand{\idty}{{\leavevmode{\rm 1\mkern -5.4mu I}}}
\newcommand{\Ir}{Z\!\!\!Z}
\newcommand{\Ibb}[1]{ {\rm I\ifmmode\mkern 
            -3.6mu\else\kern -.2em\fi#1}}
\newcommand{\ibb}[1]{\leavevmode\hbox{\kern.3em\vrule
     height 1.2ex depth -.3ex width .2pt\kern-.3em\rm#1}}
\newcommand{\Nl}{{\Ibb N}}
\newcommand{\Cx}{{\ibb C}}                                          
\newcommand{\Rl}{{\Ibb R}}                                          

\newcommand{\bi}{b_{ik}}
\newcommand{\A}{{\cal A}}
\newcommand{\U}{{\cal U}}

\newcommand{\nik}{\nabla_{ik}}
\newcommand{\nki}{\nabla_{ki}}

\newcommand{\beq}{\begin{equation}}
\newcommand{\eeq}{\end{equation}}
\begin{document} 
\begin{center}
\vspace*{1.0cm}

{\LARGE{\bf Cellular Networks as Models for\\[0.2cm] Planck-Scale Physics}} 

\vskip 1.5cm

{\large {\bf Manfred Requardt }} 

\vskip 0.5 cm 

Institut f\"ur Theoretische Physik \\ 
Universit\"at G\"ottingen \\ 
Bunsenstrasse 9 \\ 
37073 G\"ottingen \quad Germany

\end{center}

\vspace{1 cm}

\begin{abstract}
Starting from the working hypothesis that both physics and the corresponding
mathematics have to be described by means of discrete concepts on the
Planck scale, one of the many problems one has to face in this
enterprise is to find the discrete protoforms of the building blocks
of our ordinary continuum physics and mathematics. We base our own
approach on what we call `{\it cellular networks}', consisting of
cells (nodes) interacting with each other via bonds (figuring as
elementary interactions) according to a certain `{\it local
  law}'. Geometrically our dynamical networks are living on {\it
  graphs}. Hence a substantial amount of the investigation is devoted
to the developement of various versions of discrete
(functional) analysis and geometry on such (almost random)
webs. Another important topic we address is a suitable concept of {\it
  intrinsic (fractal) dimension} on erratic structures of this
kind. In the course of the investigation we make comments
concerning both different and related approaches to quantum gravity
as, say, the {\it   spin network framework}. It may perhaps be said
  that certain parts of our programme seem to be a realisation
of ideas sketched by Smolin some time ago (see the introduction).
\end{abstract} \newpage

\section{Introduction}
\noindent There exists a certain suspicion in parts of the scientific community
that nature may be discrete or rather ``behaves discretely'' on the
Planck scale. But even if one is willing to agree with this ``working
philosophy'', it is far from being evident what this vague metaphor
actually has to mean or how it should be implemented into a
concrete and systematic inquiry concerning physics and mathematics in the
Planck regime.

There are basically two overall attitudes as to ``discreteness on the
Planck scale'', the one comprising approaches which start (to a
greater or lesser degree) from continuum concepts (or more
specifically: concepts being more or less openly inspired by them) and
then try to detect or create modes of ``discrete behavior'' on very
fine scales, typically by imposing quantum theory in full or in parts
upon the model system or framework under discussion.

There are prominent and very promising candidates in this class like
e.g. {\em `string theory'} or {\em `loop quantum gravity'}. Somewhat
intermediate is a more recent version (or rather: aspect) of the
latter approach, its {\em `polymer'} respectively {\em `spin network'}
variants. As these approaches are meanwhile widely known we refrain
from citing from the vast corresponding literature. We recommend
instead as more recent reviews as to the latter approach, containing some
cursory remarks about the former together with a host of references,
\cite{Smolin} and \cite{Rovelli} and, as a beautiful introduction to the
conceptual problems of quntum gravity in general, \cite{Isham}. 

On the other hand one could adopt an even more speculative and radical
attitude and approach the Planck regime from more or less the opposite
direction by developing a framework almost from scratch which has
``discreteness'' already built in into its very building blocks and
then try to reconstruct, working, so to speak, ``bottom up'', all the
continuum concepts of ordinary space-time physics as sort of {\em
  `collective quantities'} like e.g. {\em `collective excitations'}
via the cooperation of many microscopic (discrete) degrees of freedom.
If one is very bold one could even entertain the idea that the quantum
phenomena as such are perhaps not the eternal and irreducible
principles as they are still viewed today by the majority of
physicists but, to the contrary, may emerge as derived and secondary
concepts, together with gravitation, from a more primordial truly
discrete and {\em `combinatorial'} theory. As to corresponding
strategies see also the perhaps prophetical remarks of R.Penrose in
\cite{Penrose} or the bundle of ideas presented in \cite{Smolin2}
beginning roughly at page 267. To the wider context belong also the
ideas of Sorkin, Balachandran et al. (see e.g. \cite{Sorkin}), while
the approach of 't Hooft (\cite{Hooft}) is noteworthy because of its
emphasis on the possible usefulness of {\it cellular automata}
(i.e. simpler and more rigid variants of our {\it cellular networks}).

We would like to add a perhaps clarifying remark concerning 
reference \cite{Smolin2}. The mentioned ideas are sketched almost at the end of a
long contribution and we stumbled over them only very recently
(February 1998). One could perhaps venture to say that the framework
we are going to develop in the following and in related papers (see
\cite{Req1}) to \cite{Req4} is, at least in part, an implementation of
Smolin's programme. This concerns in particular one of the main
motivations of both Smolin's and our approach, namely the hope that,
proceeding in this way, the peculiar form of nonlocality or
entanglement observed in quantum theory can actually be implemented and thus
better understood.

It goes without saying that such a radical approach is beyond the
reach of direct experimental verification in the strict sense for the
foreseeable future (as is the case with the other frameworks mentioned
above). As a substitute one rather has to rely on inner theoretical
criteria, among other things, the capability to generate the hierarchy
of increasingly complex patterns we are observing in nature or in our present day
{\em `effective theories'}, which describe the various regimes many orders of
magnitude away from the Planck scale, while introducing, on the other hand, as few simple
and elementary assumptions as possible. More specifically, one would
like such a framework to provide clues how the continuum concepts of
ordinary space-time physics/mathematics may emerge in a natural
manner from their respective {\em `discrete protoforms'}.

Another more aesthetic criterion would be a kind of natural
convergence of the different approaches towards a common substructure
which is discrete in a really primordial way. Indications for such a 
convergence can, in our view, be detected in the various lines of
research going on presently. {\em `spin networks'} or {\em `polymer
  states'} are e.g. cases in point where modes of discreteness do
emerge from a at first glance continuous environment. Furthermore it
may well be that {\em `string field theory'} will turn out to live on
a more discrete and singular substratum than presently suspected (some
speculative remarks pointing in this direction can e.g. be found at
the end of \cite{Castro}), a catchword being {\em `fractal geometry'}. A
brief but illuminating analysis concerning such a possible convergence
in the future towards a decidedly discrete and {\em `combinatorial'}
common limit has been given in section 8 of \cite{Smolin}. A bundle of
related ideas with which we sympathize is developed by Nottale (see
e.g. \cite{Nottale}). 

In the following sections we will embark on the developement of a
conceptual framework which has a pronouncedly combinatorial flavor and
makes contact with various branches of modern discrete mathematics as
e.g. {\em algebraic combinatorics}, {\em (random)graph theory} or {\em
  noncommutative geometry}. The paper is more or less divided into
three parts. The first part deals with the dynamics of `{\em cellular
  networks}' we will introduce below. Insofar our approach is
different from perhaps more canonical approaches which rely in a more
or less open way on ideas being already employed in the continuum
versions as e.g. `{\em fields}', `{\em path integrals}' or a relative
regular substratum like a periodic lattice structure. Quite to the
contrary, we regard most of these concepts as `{\em emergent
  structures}' according to our working philosophy. In other words,
these patterns need not have identifiable counterparts on the
primordial scale.

As to such questions of principle we share the philosophy of
e.g. Smolin, Ashtekar (and quite a few others, see e.g. section 6.1 in
\cite{Ashtekar}) that if one agrees that a
continuum theory has to be understood as the coarse grained limit of
an underlying `{\em critical}' discrete and more fundamental theory ,
there is no sufficient reason to surmise that the primordial building blocks
or the dynamics at perhaps the ultimate bottom of our universe will
share a lot of similarities with the (almost) continuum concepts of
our `{\em effective theories}' living in a regime of at most
intermediate energies many orders of magnitude away from the Planck
regime.

The second part of our paper consists of a presentation of a couple of
ways how one can introduce versions of {\em discrete analysis} and
{\it geometry} in such erratic
environments we are going to study in the following. At the same time
we undertake to compare our approach with other existing ones (being,
in part, derivatives of noncommutative geometry in the sense of
Connes et al).

In the last part we develop dimensional concepts on such discrete
spaces. We would like to stress the point that we consider it to be of
crucial importance not to base such concepts on anything which could
be identified as a variant of a continuum or embedding space. As to
these ideas much more can be found in \cite{Req1}, in particular
concerning various conceptual relations to {\em fractal geometry}.

We want to close this introduction with a perhaps clarifying remark.
To discretize continuum quantum gravity (or rather certain models),
e.g. to put it on a lattice or triangulate it, is of course not new,
but most of the approaches we are aware of discretize continuum
theories living in a fixed embedding space with a definite dimension
in the standard sense. Furthermore the kind of discretizations being
employed are typically relatively regular while in our approach the
underlying graphs are rather random structures with vertices and bonds
as dynamical variables which can even be switched on or off. As a
consequence the dimension of our networks is an `{\em emergent}' and
dynamical, perhaps even {\em fractal}, property of the system. To
mention only two representative paper of the many others employing or
reviewing strategies of the former type see e.g. \cite{Agishtein} or
\cite{Williams} and for more references the thesis of Bakker
(\cite{Bakker}), the lecture of T.D.Lee (\cite{Lee}) should perhaps
also be mentioned in this context. This does not of course mean that there are no
relations between these different approaches but we prefer to discuss
them elsewhere in order to keep the paper readable.

\section{The Cellular Network Environment}
In this section we will sketch the type of model systems on which the
following analysis will be based. As already said in the introduction
we will start from a rather primordial level, trying to make no
allusions whatsoever to continuum concepts. We will then show how
protoforms of ideas and notions, playing a key role in ordinary
continuum physics/mathematics emerge in a relatively natural and
unforced way from this framework. Cases in point are e.g. concepts
like {\em `dimension'}, {\em `differential structure'}, the idea of
{\em `physical points'} (being endowed with an internal structure),
the web of which establishes the substratum of macroscopic space-time,
and other geometrical/topological notions. The framework turns out to
be even rich enough to support a full fledged kind of
{\em `discrete functional analysis'}, comprising e.g. {\em `Laplace-,
  Dirac operators'} and all that. It is perhaps particularly
noteworthy that an advanced structure like {\em `Connes' spectral
  triple'} shows up in a very natural way within this context.

Besides the reconstruction of basic concepts of continuum
physics and mathematics another goal is to describe the micro dynamics
going on in this discrete substratum over (in) which macroscopic
space-time is floating as a kind of coarse grained
{\em `superstructure'}, the formation of {\em `physical points'} and
their mutual entanglement, yielding the kind of
{\em `near-/far-order'} or {\em `causal structure'} we are used to from
continuum space-time.

To this end we view this substratum as, what we like to call, a
{\em `cellular network'}, consisting of {\em `nodes'} and
{\em `bonds'}. The nodes are assumed to represent certain elementary
modules (cells or ``monads'') having a discrete, usually simple,
internal state structure, the bonds modeling elementary direct
interactions among the nodes. As an important ingredient, these bonds
are dynamical insofar as they are capable to be in a (typically limited)
number of {\em `bond states'}, thus implementing the varying strength
of the mutual interactions among the cells.

It is a further characteristic of our model class that these
interactions are not only allowed to vary in strength but, a fortiori,
can be switched off or on, depending on the state of their local 
environment. In other words, stated in physical terms, bonds can be 
created or annihilated in the course of network evolution, which
(hopefully) enables the system to undergo {\em `geometric phase
  transitions'} being accompanied by an {\em `unfolding'} and
{\em `pattern formation'}, starting e.g. from a less structured
chaotic initial phase. To put it briefly: in contrast to, say,
{\em `cellular automata'}, which are relatively rigid and regular in
their wiring and geometric structure (in particular: with the bonds
typically being non-dynamical), our cellular networks do not carry such
a rigid overall order as an external constraint (e.g. a regular
lattice structure); their ``wiring'' is dynamical and thus behaves randomly
to some extent. The clue is that order and modes of regularity are
hoped to emerge via a process of {\em `self-organization'}.
\\[0.5cm]
{\bf 2.1 Definition (Class of Cellular Networks)} \hfill  
\begin{it}
\begin{enumerate}
\item ``Geometrically'' our networks represent at each fixed
{\em `clock time'} a {\em `labeled graph'}, i.e. they consist of nodes
\{$n_i$\} and bonds \{$b_{ik}$\} or \{$d_{ik}$\} (see the next section), with the bond $b_{ik}$ connecting
the nodes (cells) $n_i$, $n_k$. We assume that the graph has neither
elementary loops nor multi-bonds, that is, only nodes with $i\neq k$
are connected by at most one bond.
\item At each site $n_i$ we have a local node state $s_i\in
q\cdot\Ir$ with $q$, for the time being, a certain not further
specified  elementary quantum. The bond variables $J_{ik}$, attached
to $b_{ik}$, are in the most simplest cases assumed to be two- or
three-valued, i.e. $J_{ik}\in \{\pm 1\}\quad\mbox{or}\quad J_{ik}\in \{\pm
  1,0\}$
\end{enumerate}
\end{it}
\vspace{0.5cm}
{\it Remarks} \hfill 
\begin{enumerate}
\item In the proper graph context the notions {\em `vertex'} and 
{\em `edge'} are perhaps more common (see e.g. \cite{Bollobas}). As to some
further concepts used in graph theory see below.
\item This is, in some sense, the simplest choice one can
make. It is an easy matter to employ instead more complicated internal
state spaces like, say, groups, manifolds etc. One could in particular
replace $\Ir$ by one of its subgroups or impose suitable  boundary 
conditions.
\item In the following section we will give the bonds $b_{ik}$ an 
{\em `orientation'}, i.e. (understood in an precise
algebraic/geometric sense) $b_{ik}=-b_{ki}$.  
This implies the compatibility conditions $J_{ik}=-J_{ki}$.
\end{enumerate}
\vspace{0.5cm}

In a next step we have to impose a dynamical law on our model
network. In doing this we are of course inspired by {\em `cellular
  automaton laws'} (see e.g. \cite{Toffoli}). The main difference is however
that in our context also the bonds are dynamical degrees of freedom
and that, a fortiori, they can become dead or alive (active or
inactive), so that the whole net is capable of performing drastic 
topological/geometrical changes in the course of clock time.

A particular type of a dynamical `{\it local law}' is now introduced
as follows: We assume that all the nodes/bonds at `{\it (clock) time}'
$t+\tau$, $\tau$ an elementary clock time step, are updated according
to a certain local rule which relates for each given node $n_i$ and
bond $b_{ik}$ their respective states at time $t+\tau$ with the states
of the nodes/bonds of a certain fixed
local neighborhood at time $t$.

It is important that, generically, such a law does not lead to a
reversible time evolution, i.e. there will typically exist attractors
in total phase space (the overall configuration space of
the node and bond states).

A crucial ingredient of our network
laws is what we would like to call a `{\it hysteresis interval}'. We
will assume that our network, called in the following $QX$ (``{\it
  quantum space})'', starts from a densely entangled `{\it
  initial phase}' $QX_0$, in which practically every pair of nodes is
on average connected by an `{\it active}' bond, i.e. $J_{ik}=\pm1$.
Our dynamical law will have a built-in mechanism which switches bonds
off (more properly: sets $J_{ik}=0$) if local fluctuations among the node
states become too large. Then there is hope that this mechanism
may trigger an `{\it unfolding phase transition}', starting from a
local seed of spontaneous large fluctuations towards a new phase (an
attractor) carrying a certain `{\it super structure}', which we would
like to relate to the hidden discrete substratum of space-time
(points).

One example of such a law is given in the following definition.
\\[0.5cm]
{\bf 2.2 Definition (Local Law)}
\begin{it}
At each clock time step a certain `{\it quantum}' $q$
is transported between, say, the nodes $n_i$, $n_k$ such that 
\beq s_i(t+\tau)-s_i(t)=q\cdot\sum_k
  J_{ki}(t)\eeq
(i.e. if $J_{ki}=+1 $ a quantum $q$ flows from $n_k$ to $n_i$ etc.)\\
The second part of the law describes the back reaction on the bonds
(and is, typically, more subtle). This is the place where the
so-called `{\it hysteresis interval}' enters the stage. We assume the
existence of two `{\it critical parameters}'
$0\leq\lambda_1\leq\lambda_2$ with:
\beq J_{ik}(t+\tau)=0\quad\mbox{if}\quad
  |s_i(t)-s_k(t)|=:|s_{ik}(t)|>\lambda_2\eeq
\beq J_{ik}(t+\tau)=\pm1\quad\mbox{if}\quad 0<\pm
  s_{ik}(t)<\lambda_1\eeq
with the special proviso that
\beq J_{ik}(t+\tau)=J_{ik}(t)\quad\mbox{if}\quad s_{ik}(t)=0
\eeq
On the other side
\beq J_{ik}(t+\tau)= \left\{\begin{array}{ll} 
\pm1 & \quad J_{ik}(t)\neq 0 \\
0    & \quad J_{ik}(t)=0
\end{array} \right. \quad\mbox{if}\quad
\lambda_1\leq\pm
  s_{ik}(t)\leq\lambda_2 
\eeq
In other words, bonds are switched off if local spatial charge
fluctuations are too large, switched on again if they are too
small, their orientation following the sign of local charge
differences, or remain inactive.
\end{it}
\\[0.5cm]
{\it Remark} \hfill 
\begin{enumerate}
\item The reason why we do not choose the ``current'' $q\cdot
J_{ik}$ proportional to the ``voltage difference'' $(s_i-s_k)$ as e.g.
in Ohm's law is that we favor a {\it non-linear} network which is
capable of {\it self-excitation} and {\it self-organization} rather than
{\it self-regulation} around a relatively uninteresting equilibrium state!
The balance between dissipation and amplification of spontaneous
fluctuations has
however to be carefully chosen (``{\it complexity at the edge of chaos}'')
\item We presently have emulated these local network laws on a computer.
It is not yet clear whether this simple network law does
already everything we are expecting. In any case, it is fascinating to observe the
enormous capability of such intelligent networks to find attractors
very rapidly, given the enormous accessible phase space
\item In the above class of laws a direct bond-bond-interaction is not
yet implemented. We are prepared to incorporate such a contribution in
a next step if it turns out to be necessary. In any case there are not
so many ways to do this in a sensible way. Stated differently, the class
of possible physically sensible interactions is
perhaps not so numerous.
\item Note that -- in contrast to e.g. Euclidean lattice field theory --
the so-called {\it `clock time'} $t$ is, for the time being, not
standing on the same footing as potential ``coordinates'' in the
network (e.g.  curves of nodes/bonds). We rather suppose that so-called
{\it `physical time'} will emerge as sort of a secondary collective
variable in the network, i.e. being different from the clock time
(while being of course functionally related to it). \label{no4}
\end{enumerate}
\vspace{0.5cm}

In our view {\it Remark $4$} is consistent with the spirit of relativity. What
Einstein was really teaching us is that there is a (dynamical)
interdependence between what we experience as space respectively time, not
that they are absolutely identical! In any case, the assumption of an
overall clock time is at the moment only made just for convenience in
order to make the model system not too complicated. If our
understanding of the complex behavior of the network dynamics
increases, this assumption may be weakened in favor of a possibly
local and/or dynamical clock frequency. A similar attitude should be
adopted concerning concepts like {\it `Lorentz-(In)Covariance'} which
we also consider as {\it `emergent'} properties. It is needless to say that
it is of tantamount importance to understand the way how these
patterns do emerge from the relatively chaotic background, questions which will
be addressed in future work.

As can be seen from the definition of the cellular network, a full
scale investigation of its behavior separates quite naturally into two
parts of both a different mathematical and physical nature. The first
one comprises its more geometric/algebraic content in form of large
static graphs and their intricate structure (at, say, arbitrary but
fixed clock time), thus neglecting the details of the internal states
of bonds and nodes, the other one conveys a more dynamical flavor,
i.e. analyzing and keeping track of the topological/geometrical change
and pattern formation in the course of clock time. Both parts
represent an intricately entangled bundle of complicated problems and
do require the developement or application of a fair amount of quite
advanced (discrete) mathematics. Therefore we prefer to concentrate in
the following on the former part; as to the latter part a certain
impression of the impending problems is given in \cite{Req4} which,
however, should be considered only as a preliminary draft.

Before we embark on studying in more detail the `{\it pregeometric}'
patterns in cellular networks we want to briefly comment on the more
general structure of evolution laws in such systems. The above one is
only one candidate from a whole class of such laws. For one, it is
quite evident that the `{\it local state spaces}' living over the
respective nodes and bonds can be chosen in a more general way. For
another, the local dynamical law can be chosen more general. \\[0.5cm]
{\bf 2.3 Definition (General Local Law on Cellular Networks)}:\\
\begin{it}
Each node
$n_i$ can be in a number of internal states $s_i\in \cal S$. Each
bond $\bi$ carries a corresponding bond state $J_{ik}\in\cal J$. Then
the following general transition law is assumed to hold:
\begin{equation} s_i(t+\tau)=ll_s(\{s_k(t)\},\{J_{kl}(t)\})
\end{equation}
\begin{equation} J_{ik}(t+\tau)=ll_J(\{s_l(t)\},\{J_{lm}(t)\})
\end{equation}
\begin{equation}
(\underline{S},\underline{J})(t+\tau)=LL((\underline{S},\underline{J})(
t))
\end{equation}
where $ll_s$, $ll_J$ are two maps (being the same over the whole
graph) from the state space of a local
neighborhood of the node or bond on the lhs 
to $\cal S, J$, yielding the
updated values of $s_i$ and $J_{ik}$. $\underline{S}$ and
$\underline{J}$ denote the global states of the nodes and bonds and
$LL$ the global law built from the local laws at each node or bond.
\end{it}
\vspace{0.5cm}

Irrespectively of the technical details of the dynamical evolution law
under discussion it should emulate the following, in our view crucial,
principles, in order to match certain fundamental requirements
concerning the capability of `{\it emergent}' and `{\it complex}'
behavior.
\begin{enumerate}
\item As is the case with, say, gauge theory or general relativity,
  our evolution law on the primordial level should encode the mutual
  interaction of two fundamental substructures, put sloppily:\\
``{\it geometry}'' acting on ``{\it matter}'' and vice versa, where in
  our context ``{\it geometry}'' is assumed to correspond in a loose
  sense with the local and/or global bond states and ``{\it matter}''
  with the structure of the node states. (We will not comment further
  on this working philosophy at the moment as it represents the basis
  of forthcoming work).
\item By the same token the above selfreferential dynamical circuit of
  mutual interactions is expected to favor a kind of `{\it undulating
  behavior}' or `{\it selfexcitation}' above a return to some
  uninteresting `{\it equilibrium state}' which is frequently found
  among systems consisting of a single component which directly acts back on
  itself. This propensity for the `{\it autonomous}' generation of
  undulation patterns is in our view an essential prerequisite for
  some form of ``{\it protoquantum behavior}'' we hope to recover on
  some coarse grained and less primordial level of the network
  dynamics.
\item In the same sense we expect the possibility of switching on and
  off of bonds to generate a kind of ``{\it protogravity}''.
\end{enumerate}

We close this section with a short aside concerning the definition of
evolution laws of `{\it spin networks}' by Markopoulou, Smolin and
Borissov (see \cite{Marko} or \cite{Boris}). As in our case there are
more or less two possibilities: treating evolution laws within an
integrated space-time formalism or regard the network as representing
space alone with the time evolution being implanted via some extra
principle ( which is the way we have chosen above). Closely connected
with this question is the developement of a suitable concept of {\it
  dimension} in this context. We start to develop our personal concept
in the last section of this paper. More information can be found in
\cite{Req1}. As the interrelation of these various approaches and
frameworks is both very interesting and presently far from obvious we
plan to compare them elsewhere.

\section{Discrete Analysis on Graphs and Networks}
In the following we will show that despite of its discreteness graphs
and networks are capable of supporting a surprisingly rich
differential and geometric structure, so that the catchword `{\it
  discrete analysis}' in the headline of this section is no
exaggeration. One can even develop a full fledged `{\it discrete
  functional analysis}' with Hilbert spaces, graph Laplacian, Dirac
operator etc. This then leads into the fascinating field of describing
geometric structures and patterns on the graph with the help of
functional analytic tools (see \cite{Req3}).

While our original approach has been developed from a somewhat
different perspective there emerged nevertheless in the course of
evolution of our framework various regions of contact and overlap with
other approaches like e.g. what is called `{\it noncommutative
  geometry}' (see e.g. \cite{Connes} or \cite{Madore}; as a brief but
concise introduction we recommend also \cite{Coque}). If a certain
part of this highly abstract machinery is applied to, say, discrete
sets, one gets a version of discrete calculus which is still a
relatively abstract scheme as long as it is not interpreted within a
concrete model theory. Such an abstract calculus has been developed by
e.g. Dimakis, Mueller-Hoissen et al. (see for example
\cite{Dimakis}). In the following we will attempt, among other things,
to relate these in some respects different, in other respects related
approaches with our own framework.

We will develop below various schemes with only the first one being
related in some respects to noncommutative geometry. The other ones
are standing on a more or less independent footing. In this first
approach context we treat the network as a static {\it labelled
  graph}, consisting solely of {\it nodes} and {\it bonds}. We
start with
some graph theoretical concepts:\\[0.5cm]
{\bf 3.1 Definition (Simple Locally Finite (Un)directed Graph)}:\\
\begin{it}
\begin{enumerate}
\item We write the `simple' graph as $G:=(V,E)$ where $V$ is the
  countable set of nodes $\{n_i\}$ (or vertices) and $E$ the set of
  bonds (edges). The graph is called simple if there do not exist
  elementary `loops' and `multiple edges', in other words: each
  existing bond connects two different nodes and there exists at most
  one bond between two nodes. (We could of course also discuss more
  general graphs). Furthermore, for simplicity, we assume the graph to
  be connected, i.e. two arbitrary nodes can be connected by a
  sequence of consecutive bonds called an `edge sequence' or `walk'. A
  minimal edge sequence, that is one with each intermediate node
  occurring only once, is called a `path' (note that these definitions
  may change from author to author).
\item We assume the graph to be `{\it locally finite}', that is, each
  node is incident with only a finite number of bonds. Sometimes it is
  useful to make the stronger assumption that this `{\it vertex
  degree}', $v_i$, (number of bonds being incident with $n_i$), is
  globally bounded away from $\infty$.
\item One can give the edges both an `{\it orientation}' and a `{\it
    direction}' (these two slightly different geometric concepts are
    frequently intermixed in the literature). In our context we adopt
    the following convention: If two nodes $n_i,n_k$ are connected by
    a bond, we interpret this as follows: There exists a `{\it
    directed bond}', $d_{ik}$, pointing from $n_i$ to $n_k$ and a {\it
    directed bond}, $d_{ki}$, pointing in the opposite direction. In
    an algebraic sense, which will become clear below, we call
    their `{\it superposition}'
\beq b_{ik}:= d_{ik}-d_{ki}= -b_{ki}\eeq
the corresponding `{\it oriented bond}' (for obvious reasons; the
    directions are fixed while the orientation can change its sign). In
    a sense the above reflects the equivalence of an `{\it undirected
    graph}' with a `{\it directed multi-graph}' having two directed
    bonds pointing in opposite directions for each undirected bond.
\end{enumerate}
\end{it}
\vspace{0.5cm}
We now take the elementary building blocks $\{n_i\}$ and $\{d_{ik}\}$
as basis elements  of a certain hierarchy of vector spaces over, say,
$\Cx$ with scalar product
\beq
<n_i|n_k>=\delta_{ik}\quad<d_{ik}|d_{lm}>=\delta_{il}\cdot\delta_{km}\eeq
{\bf 3.2 Definition (Vertex-, Edge-Space)}:
\begin{it}
 The vector spaces (or {\it
  modules}) $C_0$ and $C_1$, consist of the finite sums 
\beq f:=\sum f_in_i\quad\mbox{and}\quad g:=\sum g_{ik}d_{ik}\eeq
$f_i,g_{ik}$ ranging over a certain given {\it field} like e.g. $\Cx$
or {\it ring} like e.g. $\Ir$ in case of a module.\\[0.5cm]
\end{it}
{\it Remark}:
\begin{enumerate}
\item These spaces can be easily completed to {\it Hilbert spaces} (as
  it was done in \cite{Req3}) by assuming
\beq \sum |f_i|^2<\infty\quad \sum |g_{ik}|^2<\infty\eeq
if one chooses the field $\Cx$.
\item One can continue this row of vector spaces in a way which is
  common practice in, say, {\it algebraic topology} (for more details
  see belöow). In this context they could equally well be called `{\it
  chain complexes}'.
\item Evidently the above vector spaces could as well be viewed as
  {\it discrete function spaces} over the {\it node-, bond set} with
  $n_i,d_{ik}$ representing the elementary {\it indicator functions}.
\end{enumerate}
\vspace{0.5cm}

In the same spirit we can now introduce two linear maps between
$C_0,C_1$ called for obvious reasons `{\it boundary-}' and `{\it
  coboundary map}'. On the basis elements they act as
follows:\\[0.5cm]
{\bf 3.3 Definition/Observation ((Co)boundary Operator)}:
\begin{it}
\beq \delta:\; d_{ik}\to n_k\quad\mbox{hence}\quad b_{ik}\to
n_k-n_i\eeq
\beq d:\; n_i\to \sum_k(d_{ki}-d_{ik})=\sum_k b_{ki}\eeq
and linearly extended. That is, $\delta$ maps the directed bonds
$d_{ik}$ onto the terminal node  and $b_{ik}$ onto its `boundary',
while $d$ maps the node $n_i$ onto the sum of the `ingoing' directed
bonds minus the sum of the `outgoing' directed bonds or on th sum of
`oriented' ingoing bonds $b_{ki}$.
\end{it}
\vspace{0.5cm}

The following results show, that these definitions lead in fact to a
kind of `{\it discrete differential calculus}'.\\[0.5cm]
{\bf 3.4 Observation (Discrete Differential Forms)}:
\begin{it}
From the above it follows that
\beq df=d(\sum f_in_i)=\sum_{k,i}(f_k-f_i)d_{ik}\eeq
\end{it}
\vspace{0.5cm}
One could now enter the field of true {\it discrete functional
  analysis} by defining the so-called `{\it graph Laplacian}'.\\[0.5cm]
{\bf 3.5 Definition/Observation (Graph Laplacian)}:
\begin{it}
\beq \delta df=-\sum_i(\sum_k f_k-v_i\cdot
f_i)n_i=-\sum_i(\sum_k(f_k-f_i))n_i=:-\Delta f \eeq
where $v_i$ denotes the node degree or `valency' defined above and the
$k$-sum extends over the nodes adjacent to $n_i$.
\end{it}
\\[0.5cm]
As to more results along these lines of reasoning see \cite{Req3}. In
the following we would however prefer to develop a framework which is
more in the spirit of `{\it discrete geometry}' or abstract `{\it
  combinatorial topology}'.

\subsection{The Graded Semi-Differential Algebra of Strings or Walks}
The elementary algebraic/geometric building blocks of our framework
will be the nodes and (un)directed bonds. In a next step we use them to form
`{\it edge sequences (walks)}' or `{\it strings}'.\\[0.5cm]
{\bf 3.6 Definition/Observation (Admissible Strings)}:
\begin{it}
With $C_0,C_1$ denoting the vector spaces spanned by the nodes and
bonds, $C_k$ comprises finite sums of admissible edge sequences
consisting of $k$ consecutive edges or $(k+1)$ consecutive nodes. We
write them as
\beq d_{i_0i_1}\cdot d_{i_1i_2}\cdots
d_{i_{k-1}i_k}\quad\mbox{or}\quad n_{i_0}\ldots n_{i_k}\eeq
where by admissible we mean that each pair of consecutive nodes is
connected by a bond. The multiplication sign standing between the
respective edges is explained below and the above strings are now
interpreted as normalized basis elements in the vector space $C_k$
over, say, $\Cx$.\end{it}\\[0.5cm]
{\it Remark}: Note that the repeated occurrence of a particular node
or bond in such a string is not forbidden, whereas two consecutive
nodes are always different. A fortiori a substring of the form
$d_{ik}\cdot d_{ki}$ or $n_in_kn_i$ is admissible. Such strings or
substrings are algebraically useful if one wants to have the notion of
`{\it inverse string}' and a corresponding multiplication structure
(thinking of more general algebraic concepts like e.g. groupoids
etc.).
\vspace{0.5cm}

We can now extend the (co)boundary maps $\delta,d$ to consecutive
pairs of spaces, $C_k,C_{k+1}$ by combining geometric imagery with
experience from algebraic topology.\\[0.5cm]
{\bf 3.7 Definition/Observation}:
\begin{it}
With a certain admissible string $n_0\ldots n_k$ given (for notational
simplicity we write $k$ instead of $i_k$) we define $\delta,d$ as
follows:
\beq \delta (n_o\ldots n_k):=\sum_{i=0}^k
(-1)^in_0\ldots\widehat{n_i}\ldots n_k\eeq
where the hat means that the respective node has to be deleted with
the proviso that the string is mapped to zero if $n_{i-1},n_{i+1}$ are
not connected by a bond, i.e. if the new string is not admissible.

The extension of the coboundary operator runs as follows:
\beq d(n_0\ldots n_k):=\sum_{n_{\nu},i}(-1)^in_0\ldots
n_{i-1}n_{\nu}n_i\ldots n_k \eeq
where $i$ runs from $0$ to $(k+1)$ and admissible nodes are inserted
before the node $n_i$ or for $i=(k+1)$ after the last node $n_k$. The
insertion runs over all nodes being connected with both $n_{i-1}$ and
$n_i$.\end{it}\\[0.5cm]
The geometric picture behind this scheme is quite transparent. A given
string in $C_k$ is mapped onto a superposition of strings in $C_{k-1}$
or $C_{k+1}$ with appropriate weights as prefactors.

After these preparatory steps we can now set up a certain discrete
differential structure with a strong geometric flavor. We have the
$\Nl_0$-graded vector space 
\beq C:=\sum C_k\quad k\in\Nl\cup\{0\}\eeq
with the `{\it raising}' and `{\it lowering}' operators $d$ and
$\delta$, mapping the `{\it level sets}' of homogeneous elements,
$C_k$, into each other.

In a first step we make $C$ into a `{\it graded algebra}':\\[0.5cm]
{\bf 3.8 Definition/Observation (Graded Algebra)}:
\begin{it}
We define multiplication of homogeneous elements via `{\it
  concatenation}', i.e:
\beq (n_{i_0}\ldots n_{i_k})\cdot(n_{j_0}\ldots
  n_{j_l})=(n_{i_0}\ldots(n_{i_k}\cdot n_{j_0})\ldots n_{j_l})\in
  C_{k+l}\eeq
where $n_{i_k}\cdot n_{j_0}$ means the natural algebra product
  structure on $C_0$, i.e. pointwise multiplication
\beq f\cdot g=(\sum f_in_i)\cdot(\sum g_in_i)=\sum f_ig_in_i\eeq
that is $n_{i_k}\cdot n_{j_0}=\delta_{i_kj_0}n_{j_0}$.
\end{it}\\[0.5cm]
Geometrically this describes the glueing together of strings or walks,
  the product being zero if the end node of the first string is
  different from the initial node of the second string (in other
  words, this algebra has a lot of `{\it zero divisors}').

As a consequence $C$ has a natural `{\it bimodule structure}' over
$C_0$.\\[0.5cm]
{\bf 3.9 Corollary (Bimodule Structure)}:
\begin{it}
With\end{it} 
\beq n_i\cdot(n_i\ldots n_k)=(n_i\cdot n_1)n_2\ldots
n_k=\delta_{i1}\cdot n_1\ldots n_k\eeq
\beq (n_1\ldots n_k)\cdot n_i=n_1\ldots(n_k\cdot n_i)=\delta_{ki}\cdot
(n_1\ldots n_k)\eeq
\begin{it}and linear extension $C$ becomes a bimodule over the algebra
$C_0$. Algebraically $n_1,n_k$ play the role of `{\it left-}' and
`{\it right-}'identities with respect to the above string.
\end{it}
\vspace{0.5cm}

In order to exhibit the connection to differential calculus in the
style of Connes or Dimakis et al., we make the following observation
concerning the interplay of algebraic and differential
structure.\\[0.5cm]
{\bf 3.10 Observation}:
\begin{it}
Given an admissible string $n_0\ldots n_k$ and applying the operator
$d$ on each of the nodes $n_1,\ldots,n_k$ we have the identities:\end{it}
\beq n_in_{i+1}=n_i\cdot dn_{i+1}=dn_i\cdot n_{i+1}\eeq
\begin{it} and hence\end{it}
\beq n_0\ldots n_k=(n_0n_1)\cdot(n_1n_2)\cdots(n_{k-1}n_k)\eeq
\begin{it}(concatenation of bonds, i.e. elementary strings)\end{it}
\beq =(n_0dn_1)\cdot(n_1dn_2)\cdots(n_{k-1}dn_k)=n_0\cdot dn_1\cdot
dn_2\cdots dn_k=:n_0dn_1\ldots dn_k\eeq
Proof: The latter manipulations exploit the algebraic (concatenation)
rules we have already established above (remember the definition of
$dn_i$ as a certain superposition of $1$-strings). The former
identities hold because
\beq n_i\cdot d_{i+1}=n_i\cdot(\sum_k (d_{k(i+1)}-d_{(i+1)k}))\eeq
with $d_{k(i+1)}$ denoting the string $n_kn_{i+1}$ etc., hence
$n_i\cdot d_{k(i+1)}=\delta_{ik}\cdot d_{i(i+1)}$. That is, it
survives only the one term $d_{i(i+1)}$ or $n_in_{i+1}$ since the
other term $n_i\cdot d_{(i+1)k}$ is always zero by the definition of
the concatenation product.\hfill$\Box$
\\[0.5cm]
It is exactly these abstract (usually uninterpreted) objects
$a_0da_1\ldots da_k$ which occur in the abstract formalism of, say,
`{\it non-commutative de Rham complexes}' or the `{\it universal
  differential algebra}' (see e.g. \cite{Connes} or for the case of
discrete sets \cite{Dimakis}).\\[0.5cm]
{\it Remarks}
\begin{enumerate}
\item In an earlier version (\cite{Req5}) we based most of the
  algebraic calculus on the concept and properties of tensor products
  (of e.g. modules) which, we think, has obscured a little bit the very
  suggestive geometric imagery. On the other side it is perhaps closer
  to the abstract framework of non-commutative geometry.
\item In the same paper we showed that our strings realize also the
  abstract structure of a `{\it groupoid}'. For sake of brevity we
  omit the developement of the corresponding framework in this paper.
\end{enumerate}
\vspace{0.5cm}

The above shows that our strings form a graded algebra with
concatenation as multiplication. In a next step we show that the
coboundary operator $d$ fulfills the so-called (graded) `{\it Leibniz
  rule}' (wellknown from e.g. the exterior differential
algebra)\\[0.5cm]
{\bf 3.11 Observation (Graded Leibniz Rule)}:{\it With $w_1,w_2$ two
  strings, it holds}
\beq d(w_1\cdot w_2)=dw_1\cdot w_2+(-1)^{deg(w_1)}w_1\cdot dw_2\eeq
{\it where $deg(w_1)=$ number of edges in $w_1$\\
In the special case of zero-strings or functions from $C_0$ the
analogous formula reads:}
\beq d(f\cdot g)=df\cdot g+f\cdot dg\eeq
{\it as $deg(f)=0$}.\\[0.5cm]
Proof: Whereas the latter result is contained in the former one, we
prefer to supply an extra proof as kind of a warm up.
\beq d(f\cdot g)=\sum_i f_ig_idn_i=\sum_i(\sum_k (n_kn_i-n_in_k))\eeq
\beq df\cdot g+f\cdot dg=(\sum_{i,k} f_ig_i(n_kn_i)-\sum_{i,k}
f_ig_k(n_in_k))+(\sum_{i,k} f_kg_i(n_kn_i)-\sum_{i,k}
f_ig_i(n_in_k))\eeq
The mixed $f_ig_k$-terms in the latter equation cancel each other
(after a change of dummy variables), hence the result.

In the case of general strings with 
\beq w_1=n_{i_0}\ldots n_{i_k},w_2=n_{j_0}\ldots n_{j_l}\eeq
we can split the action of $d$ on $w_1\cdot w_2$ in two groups of
terms, the one comprising the action {\it within} $w_1$ or $w_2$, the
other consisting of the terms which arise from the action of $d$ on
the ''{\it interface}'' between $w_1$ and $w_2$.

The terms in the first group occur also in $dw_1\cdot w_2$ if $d$ acts
within $w_1$ or in $w_1\cdot dw_2$ modified by an extra weight factor
$(-1)^{deg(w_1)}$ if $d$ acts within $w_2$.

As to the interface-terms, there are three possibilities:
\begin{enumerate}
\item $n_{i_k}$ and $n_{j_0}$ are no nearest neighbors, i.e. they are
  more than one bond distance apart. In that case $w_1\cdot w_2=0$,
  hence $d(w_1\cdot w_2)=0$. On the other side, under this proviso no
  term in $dw_1$ is incident with $w_2$ and vice versa, that is
  $dw_1\cdot w_2=0=w_1\cdot dw_2$.
\item $n_{i_k}\neq n_{j_0}$ but they are nearest neighbors. This
  implies $w_1\cdot w_2=0=d(w_1\cdot w_2)$. Now there is exactly one
  term in $dw_1$ incident with $w_2$, namely $n_{i_0}\ldots
  n_{i_k}n_{j_0}$ carrying the weight $(-1)^{i_k+1}$. A corresponding
  term  occurs in $dw_2$, carrying the weight $(-1)^{0}=1$. In the
  graded Leibniz rule this latter term (standing in front of $w_1\cdot
  dw_2$) is multiplied by $(-1)^{i_k}$, yielding the superposition
\beq (-1)^{i_k}((-1)^1\cdot(string)+(-1)^0\cdot(string))=0\eeq
\item The remaining possibility, $n_{i_k}=n_{j_0}$ is very simple. The
  corresponding end term in $dw_1$ leads away from
  $n_{i_k}=n_{j_0}$. So its concatenation with $w_2$ is zero. The same
  holds for the initial term in $dw_2$. That is, in this situation all
  the non-zero terms in $dw_1\cdot w_2,w_1\cdot dw_2$ are internal
  terms, falling in the first group, discussed above and for which the
  stated identit6y has already been verified.
\end{enumerate}
This proves the graded Leibniz rule for strings (a different proof
have been given in \cite{Nowotny}).\hfill$\Box$
\vspace{0.5cm}

We have now to explain why we call our algebra a `{\it
  semi-differential algebra}'. To this end we show that in general,
  i.e. if the underlying graph is not `{\it complete}' or a `{\it
  simplex}', the relations $d\cdot d=0$ and $\delta\cdot\delta=0$ do
  not hold for every string or node. On the other side they are
  fulfilled for a complete graph. Phenomena like these are analyzed in
  more detail (among other things) by T.Nowotny in his Diploma
  thesis (\cite{Nowotny}) and are not difficult to prove anyhow in
  full generality. Therefore we content ourselves with giving the main
  clue of the reasoning (the phenomenon was already discussed in
  \cite{Req5}, section 4.3).

Take e.g. a graph, consisting of the nodes $n_0,n_1,n_2$ and the bonds
$n_0n_1,n_1n_2$ (plus the ``opposite bonds'' $n_1n_0,n_2n_1$). We have
\beq d(n_0)=-n_0n_1+n_1n_0\quad\mbox{and}\quad
dd(n_0)=-n_0n_1n_2+n_2n_1n_0\neq 0\eeq
On the other hand, if the bonds $n_0n_2,n_2n_0$ were present we would
have
\beq dn_0=-n_0n_1-n_0n_2+n_2n_0+n_1n_0\eeq
and it is easy to see that all the terms in $ddn_0$ do in fact
cancel.

Analyzing the general case a little bit more systematically we observe
the following: In the `{\it reduced graph}' (i.e. some bonds missing) the following can happen. Apply $d$ to a given
string which yields e.g. an insertion between node $n_{i-1}$ and node
$n_{i}$, e.g. \quad$\ldots n_{i-1}n_{\nu}n_{i}\ldots$\quad with a weight
$(-1)^i$. Applying $d$ again may yield another admissible insertion of
the type \quad$\ldots n_{i-1}n_{\mu}n_{\nu}n_{i}\ldots$\quad coming with the weight
$(-1)^i\cdot (-1)^i$ provided that $n_{\mu}$ is connected with $n_i$
and $n_{\nu}$.\\
On the other hand the ``counterterm'' with the weight
$(-1)^i\cdot(-1)^{i+1}$ may be missing as it can happen that $n_{\mu}$
is connected with $n_{i-1}$ and $n_{\nu}$ but {\it not} with $n_{i-1}$ and
$n_{i}$ so that $\ldots n_{i-1}n_{\mu}n_{i}\ldots$ does not show up in
the first step in contrast to the analogous term with $n_{\nu}$.
A similar result holds for $\delta$, hence:\\[0.5cm]
{\bf 3.12 Conclusion}:{\it We call the above algebra a
  `semi-differential algebra' since in general (if some bonds are
  missing) $dd\neq 0,\delta\delta\neq 0$ on certain strings. If, on
  the other side, the underlying graph is complete, we have $dd=\delta\delta=0$ }
\vspace{0.5cm}

These last observations are perhaps of some help if one wants to compare our
(more ``bottom up'') approach with a to some extent alternative (and
more ``top down'') approach, which starts from the concept of the
`{\it universal differential algebra}' (see the remarks at the
beginning of this section). It is then an almost trivial observation
(corresponding results being almost ubiquituous in abstract algebra)
that every `{\it differential algebra}' is the homomorphic image of
the universal one; put differently: it is isomorphic to the universal
algebra divided by a certain `{\it (differential) ideal}' (see below).
These results have then been aplied to discrete sets by
e.g. Dimakis,Mueller-Hoissen et al (\cite{Dimakis}).

On the other hand we want to stress that models, in our case networks
or graphs, tend to have a lot more interesting geometric or algebraic
structure than suggested by such an abstract viewpoint and should be
analyzed for their own sake (see also the following subsections). We
want now to briefly comment on one particular feature one should at
least be aware of if one favors the abstract approach. We would like
to call it the `{\it problem of unnatural relations}'. 

To conform with the more abstract notation adopted in the above
mentioned literature (or \cite{Req5}), we now denote the universal
differential algebra over a given set of nodes (i.e. the complete
graph) by $\Omega^u=\sum \Omega_k^u$, the reduced algebra (i.e. the
actually given graph with some bonds missing in general) by
$\Omega=\sum \Omega_k$. Furthermore, $d_u$ is the differential on
$\Omega^u$ with $d_u\cdot d_u=0$. Note that in $\Omega_k^u$ each
sequence of $(k+1)$ nodes is an admissible string.

We can now define a projector $\Pi$ which projects $\Omega^u$ onto
$\Omega$ by:
\begin{equation}\Pi(n_0\ldots n_k)=0\end{equation}
if $(n_0,\ldots ,n_k)$ is not admissible in $\Omega$
\begin{equation}\Pi(n_0\ldots n_k)=n_0\ldots n_k\end{equation}
if $(n_0,\ldots ,n_k)$ is admissible.\\[0.5cm]
{\bf 3.13 Consequence}: {\it We have} 
\begin{equation}\Pi=\Pi^2\; , \;
  \Omega^u=\Pi\Omega^u+(\idty-\Pi)\Omega^u\end{equation}
{\it with} $\Pi\Omega^u=\Omega$.{\it We could now continue by defining}
\begin{equation}d:=\Pi\circ d_u\circ\Pi\end{equation}
{\it which leaves $\Omega$ invariant but in general $d\cdot d\neq 0$ in
contrast to $d_u\cdot d_u=0$ as we have seen above}.\vspace{0.5cm}

We observe that $Ker(\Pi)$ is a two-sided
ideal $I$ consisting of the elements $n_{0\ldots k}$ having at least
one pair of consecutive nodes {\it not} being connected by a
bond in the reduced graph. However, this ideal $I$ is {\it not} left 
invariant under the action of $d_u$!
A closer analysis shows that $d_u(n_{0\ldots k})\notin I$ if $d_u$ creates
{\it 'insertions'} between non-connected neighbors in the reduced graph
s.t. non-admissible elements become admissible,
i.e. connected. We hence have:
\\[0.5cm]
{\bf 3.14 Observation}: {\it In general there exist elements $n_{0\ldots
  k}\in Ker(\Pi)$ s.t. $\Pi(d_u(n_{0\ldots k}))\neq 0$, in other
words, $I$ is in general not left invariant by $d_u$.\\
If one wants to make $\Omega$ into a real differential algebra one has to
enlarge $I$!}\\[0.5cm]
{\bf 3.15 Consequence}: {\it The ideal $I'=I+d_u\circ I$ is invariant under
$d_u$ and $d$ defines a differential algebra on the smaller algebra
$\Omega^u/I'\subset \Omega$ with $\Omega=\Omega^u/Ker(\Pi)$.\\[0.5cm]
(That $I'$ is an ideal left invariant by $d_u$ is easy to prove with
the help of the property $d_u\cdot d_u=0$)}.\vspace{0.5cm}

So much so good, but in our view there now emerges a certain
problem. $\Omega^u/I'$ is the algebra one automatically arrives at if
one defines the homomorphism $\Phi$ from $\Omega^u$ to the reduced
differential algebra in the following canonical way:
\begin{equation}\Phi:\; n_i\to n_i\quad d_un_i\to dn_i\end{equation}
i.e. under the premise that $d$ defines already another differential
algebra. It is in this restricted sense the above mentioned general
result has to be understood.

On the other hand this may (and in general: will) lead to a host of,
at least in our view,
very unnatural relations in concrete models like e.g. our network,
models which
may already carry a certain physically motivated interpretation going
beyond being a mere example of an abstract differential algebra. Note
e.g. that in our algebra $\Omega$ an element like $n_{123}$ is
admissible (i.e. non-zero) if $n_1,n_2$ and $n_2,n_3$ are
connected. $n_{123}$ may however arise from a differentiation process
(i.e. from an insertion) like $d_u(n_{13})$ with $n_1,n_3$ {\it not} connected.

This is exactly the situation discussed above:
\begin{equation}n_{13}\in I \quad\mbox{but}\quad d_u(n_{13})\notin
  I\end{equation}
Dividing now by $I'$ maps $d_u(n_{13})$ onto zero whereas there may be
little physical or geometric reason for $n_{123}$ or a certain
combination of such admissible elements being zero in our
network.\\[0.5cm]
{\bf 3.16 Conclusion}: {\it Given a concrete physical network $\Omega$ one
has basically two choices. Either one makes it into a full-fledged
differential algebra by imposing further relations which may however
be unnatural from a physical point of view and very cumbersome for
complicated networks. This was the strategy
e.g. followed in \cite{Dimakis}.\\
Or one considers $\Omega$ as the fundamental object and each
admissible element in it being non-zero. As a consequence the
corresponding algebraic/differential structure on $\Omega$ may be less
smooth at first glance ($dd\neq 0$ in general), but on the other side
considerably more natural! At the moment we refrain from making a
general judgement about the respective advantages of these two
different approaches whereas we would probably prefer the latter one}.

\subsection{The Graph as a (Higher) Simplicial Complex}
In the following we will change our geometric point of view to some
extent. In the previous subsection the geometric building blocks have
been strings or walks, that is, more or less one-dimensional objects.
Now we will be concerned with in some sense higher dimensional
patterns. To this end we start from a fixed graph $G$ with vertex set
$V$ and edge set $E$. In the literature graphs are frequently
considered as so-called `{\it one-dimensional simplicial complexes}'
with the nodes as 0-simplices and the bonds as 1-simplices. We think
this point of view is unnecessarily restrictive (while there exist of
course certain mathematical reasons for this restriction we do not
mention). Having our own working philosophy of networks as models for
microscopic space-time in mind,
this concept can and should be generalized.\\[0.5cm]
{\bf 3.17 Definition (Subsimplices)}:{\it We call a subset of $(k+1)$
  vertices a `k-simplex', $s_k$, or `complete subgraph' if every two
  nodes of $s_k$ are connected by a bond}.\\[0.5cm]
{\bf 3.18 Observation}: {\it If the node degree, $v_i$, of $n_i$ is
  smaller than $\infty$ then $n_i$ can lie in at most a finite set of
  different simplices, an upper bound being provided by the number of
  different subsets of bonds emerging from $n_i$}\\[0.5cm]
Proof: Assume that $s_k,s_l$ are two different simplices containing
$n_i$. By definition $n_i$ is linked with all the other nodes in $s_k$
or $s_l$. As these sets are different by assumption, the corresponding
subsets of bonds emerging from $n_i$ are different. On the other side,
not every subset of such bonds corresponds to a simplex (there
respective endpoints need not form a simplex), which proves the upper
bound stated above.\hfill$\Box$.\\[0.5cm]
{\bf 3.19 Consequence}:
\begin{enumerate}
\begin{it}
\item The set of subsimplices is evidently `partially ordered' by
  inclusion
\item Furthermore, if s is a simplex, each of its subsets is again a
  simplex (called a `face')
\item It follows from Observation 3.18 that each of these `chains' of
  linearly ordered simplices is finite with the same upper bound as in
  Observation 3.18. In other words each chain has a maximal element, a
  so-called `maximal subsimplex', abbreviated by mss. By the same token
  each node lies in at least one (but generically several) mss
\item Such a mss with $n_i$ being a member can comprise at most $(v_i+1)$
  nodes, in other words, its cardinality is the minimum of these
  numbers when $n_i$ varies over the mss.
\end{it}
\end{enumerate}
\vspace{0.5cm}
{\it Remark}: Such $mss$ are called in combinatorics or graph theory
`{\it cliques}'. Their potential physical role as building blocks of the
micro-structure of space-time in our particular approach has been
discussed in greater depth in \cite{Req4}.\\[0.5cm]
{\bf 3.20 Observation}: {\it The class of simplices, in particular
  the mss, containing a certain fixed node, $n_i$, can be generated in
  a completely algorithmic way, starting from $n_i$. The first level
  consists of the bonds with $n_i$ an end node, the second level
  comprises the triples of nodes (`triangles'), $(n_in_kn_l)$, with
  the nodes linked with each other and so forth. Each level set can be
  constructed from the preceding one and the process stops when a mss
  is reached}.\\[0.5cm]
{\it Remark}: Note that at each intermediate step, i.e. having
already constructed a certain particular subsimplex, one has in general
several possibilities to proceed. On the other hand, a chain of such
choices may differ at certain places from another one but may lead in
the end to the same simplex (being simply a permutation of the nodes of the
latter simplex).\vspace{0.5cm}

In a next step we can give the simplices an `{\it
  orientation}'.\\[0.5cm]
{\bf 3.21 Definition (Orientation)}: {\it The above simplices can be
  oriented via}
\beq n_{\pi(0)}\ldots n_{\pi(k)}=\mbox{sgn}(\pi)\cdot n_0\ldots n_k\eeq
{\it $\pi$ being a permutation of $(0,\ldots,k)$}\\[0.5cm]
This definition can be used to give the set of simplices an
algebraic structure by identifying the simplices which fall in the
same equivalence class with respect to $\mbox{sgn}\pi$. The respective
equivalence class will henceforth be denoted by $(n_{i_0}\ldots
n_{i_k})$ and we make these classes into an orthonormal basis of some
vector space of, say, finite sums over e.g. $\Cx$. The other class
(negative orientation) is then simply the opposite vector
$-(n_{i_0}\ldots n_{i_k})$.\\[0.5cm]
{\bf 3.22 Consequence}: {\it The above observations show that our set of
oriented simplices establish what is called in algebraic topology an `abstract'
or `combinatorial' `simplicial complex'. If the node degree is
uniformly bounded away from $\infty$ on $V$, this simplicial complex
has a `finite degree' (which is the cardinality of the largest
occurring mss)}.\\[0.5cm]
{\it Remark}: By means of an elegant argument one can show that such a
simplicial complex with degree, say, $n$ can always be embedded in $\Rl^{2n+1}$
(see e.g. \cite{Stillwell}).
\vspace{0.5cm}

On this simplicial complex we can now define two operators $\delta,d$
which, however, differ from the ones defined in subsection 3.1. We will
show in particular that $\delta\cdot\delta=d\cdot d=0$ always
holds.\\[0.5cm]
{\bf 3.23 Definition/Observation}: {\it The `boundary operator'
  $\delta$ acts in the following way on the basis elements:}
\beq \delta(n_0\ldots n_k):=\sum_{l=0}^k (-1)^l\cdot
(n_0\ldots\widehat{n_l}\ldots n_k)\eeq
{\it with $\widehat{n_l}$ being omitted as usual. Note that the terms
  on the rhs are again simplices (faces). It is a standard procedure
  to prove that $\delta\cdot\delta=0$ holds. The `coboundary operator'
  $d$ is defined via:}
\beq d(n_0\ldots n_k):=\sum_{n_{\nu}} (n_{\nu}n_0\ldots n_k)\eeq
{\it the sum on the rhs running over the nodes $n_{\nu}$ so that the
  terms under the sum form again $(k+1)$-simplices (or, put
  differently, the sum extends over all nodes with terms set to zero
  if they do not form $(k+1)$-simplices). It holds $d\cdot
  d=0$}.\\[0.5cm]
{\it Remark}: Note that in the above formula it is important to work
with equivalence classes. Otherwise the definition would become quite
cumbersome. As in the case of `{\it exterior algebra}' of e.g. {\it
  forms} the ordering in $(\ldots)$ is `{\it anticommutative}', that
is, 
\beq (n_{\nu}n_0\ldots n_k)=-(n_0n_{\nu}\ldots n_k)\eeq
Proof: We show that $d\cdot d=0$ in fact holds. We have:
\beq dd(s)=\sum_{\mu}\sum_{\nu} (n_{\mu}n_{\nu}s)\eeq
with $s$ a certain simplex. As both $(n_{\nu}n_{\mu}s)$ and $(n_{\mu}n_{\nu}s)=-(n_{\nu}n_{\mu}s)$ occur in the double sum all the
terms on the rhs cancel.\hfill$\Box$ \\[0.5cm]
{\bf 3.24 Observation}:{\it If we consider the $s_k$ as basis elements of
the vector space introduced above, it follows that $d$ is the adjoint
of $\delta$.}\\[0.5cm]
Proof: 
\beq <s_k|\delta
s'_{k+1}>=\sum_{\nu=0}^{k+1}(-1)^{\nu}\cdot<s_k|(n'_0\ldots\widehat{n'_{\nu}}\ldots
n'_{k+1})>\eeq
We see that at most one non-zero term can occur on the rhs, say,
$(-1)^{\nu}<s_k|\pm s_k>$, $\pm$ depending on the respective
orientation. On the other side, 
\beq <ds_k|s'_{k+1}>=\sum_{n_{\nu}}<(n_{\nu}n_0\ldots n_k)|(n'_0\ldots
n'_{k+1})>\eeq
Again there exists at most one non-zero term on the rhs with the node
sequence on the left side of the scalar product being a certain
permutation of the node sequence on the right side. In a first step we
permute the $n_0\ldots n_k$ on the left side so that they are standing
in the same relative order as on the right side. This yields the same
prefactor $\pm1$ as in the formula above. We then commute the
additional node $n_{\nu}$ until it occupies the same position as in
$s'_{k+1}$ yielding another prefactor
$(-1)^{\nu}$\hfill$\Box$\\[0.5cm]
{\it Remark}: This latter version is perhaps closer to the original
point of view adopted in {\it cohomology theory} (in the sense of, say, Alexander), where
cotheory typically lives on `{\it dual spaces}' (cf. e.g. \cite{Spanier}).
\\[0.5cm]
{\bf 3.25 Corollary}: {\it The above notion of orientation shows that
  e.g. the 1-simplices (i.e. bonds) correspond rather to the oriented
  bonds $b_{ik}=n_in_k-n_kn_i$ than to the directed bonds
  $d_{ik}=n_in_k$ introduced in section 3.1 (see e.g. item 3 of
  Definition 3.1). The definition of d, given in this subsection, can then be
  rewritten as}
\beq d(n_0)=\sum_{n_k}(n_kn_0)=\sum_k b_{k0}\eeq
{\it hence turns out to be consistent with the previous one given in
  subsection 3.1 at least on the first level.

On the higher levels the correspondence is slightly trickier, see the
following remark}.\\[0.5cm]
{\it Remark}: For higher simplices the correspondence is slightly more
involved. Take e.g. a {\it triangle} $(n_0n_1n_2)$. In the spirit of
subsection 3.1 one can take all the possible 3-strings which can be
built over this simplex, in other words, all the possible permutations
of $n_0n_1n_2$ and define the algebraic, oriented simplex
$(n_0n_1n_2)$ of subsection 3.2 as follows:
\beq (n_0n_1n_2):=\sum_{per}\mbox{sgn}(per)\cdot
n_{i_0}n_{i_1}n_{i_2}\eeq
with the sum running over all the permutations of $(0,1,2)$ and the
terms on the rhs being {\it strings} in the sense of subsection
3.1. It follows directly that the sum is anticommutative with respect
to rearrangements of the nodes (it may be convenient to add a
global combinatorial factor $(k+1)!^{-1}$ on the rhs ). From this point
of view the correspondence is reminiscent of the way exterior forms
are built from tensors.
\subsection{Yet another version of Discrete Calculus}
In this last subsection we briefly comment on yet another version of
discrete calculus, which is, at first glance, perhaps the most natural one and which is
absent in approaches starting from the abstract universal differential
algebra. As this variant is developed further in the diploma thesis of
T.Novotny (some results can already be found in \cite{Req4}), in
particular concerning the method of `{\it discrete
  Euler-Lagrange-Variation}', we will not elaborate very much on this point. 
\\[0.5cm]
{\bf 3.26 Definition (Partial Forward Derivative at Node (i))}:
\begin{equation} \nabla_{ik}f(i):=f(k)-f(i) \end{equation}
{\it where $n_i,n_k$ are 'nearest-neighbor-nodes', i.e. being
connected by a bond} .\\[0.5cm]
{\bf 3.27 Observation}: 
\begin{eqnarray}
\nabla_{ik}(f\cdot g)(i) & = & (f\cdot g)(k)-(f\cdot g)(i)
\nonumber\\
                         & = & \nabla_{ik}f(i)\cdot
g(i)+f(k)\cdot\nabla_{ik}g(i)\\
                         & = &
\nabla_{ik}f(i)g(i)+f(i)\nabla_{ik}g(i)+\nabla_{ik}f(i)\nabla_{ik}g(i)
\end{eqnarray}
In other words the "derivation" $\nabla$ does {\bf not} obey the
ordinary Leibniz rule. In fact, application of $\nabla$ to, say, higher powers
of $f$ becomes increasingly cumbersome (nevertheless there is a certain
systematic in it). One gets for example (with $q:=\nabla_{ik}$):
\begin{equation}q(f_1\cdots f_n)=\sum_i f_1\cdots q(f_i)\cdots
  f_n+\sum_{ij} f_1\cdots q(f_i)\cdots q(f_j)\cdots f_n+\ldots+
  q(f_1)\cdots q(f_n) \end{equation} 
Due to the discreteness of the formalism and, as a consequence, the
inevitable bilocality of the derivative there is no chance to get
something like a true Leibniz rule on this level.\vspace{0.5cm}

It is perhaps worth mentioning that the above establishes an
interesting abstract algebraic multiplication structure which is
called in \cite{Coque} a `{\it Cuntz algebra structure}' (occurring
there however in another context).\\[0.5cm]
{\bf 3.28 Observation (Cuntz algebra)}:
\begin{equation}q(f\cdot g)=q(f)\cdot g+f\cdot q(g)+q(f)\cdot
  q(g)\end{equation}
{\it and analogously for vector fields $\sum a_{ik}\nik$.\\
With $u:=1+q$ we furthermore get}:
\begin{equation}u(f\cdot g)=u(f)\cdot u(g) \end{equation}
{\it and}
\begin{equation}q(f\cdot g)=q(f)\cdot g+u(f)\cdot q(g)\end{equation}
{\it i.e. a `twisted derivation' with $u$ an endomorphism on some
  algebra, $\A$, of functions on V }.\vspace{0.5cm}

As is the case with $\nik$, the product rule for higher products can be inferred
inductively:
\begin{equation}q(f_1\cdots f_n)=\sum_i f_1\cdots q(f_i)\cdots
  f_n+\sum_{ij} f_1\cdots q(f_i)\cdots q(f_j)\cdots f_n+\ldots+
  q(f_1)\cdots q(f_n) \end{equation}
{\bf 3.29 Definition ((Co)Tangential Space)}: {\it We call the space
spanned by the $\nabla_{ik}$ at node $n_i$ the tangential space
$T_i$. Correspondingly we introduce  the space spanned by the
$d_{ik}$ at node $n_i$ and call it
the cotangential space $T_i^{\ast}$ with the $d_{ik}$ acting as linear
forms over $T_i$ via}:
\begin{equation} <d_{ik}|\nabla_{ij}>=\delta_{kj} \end{equation}\\[0.5cm]
{\bf 3.30 Definition/Observation}: {\it Higher tensor products of differential forms
at a node $n_i$ can now be defined as multilinear forms}:
\begin{equation} <d_{ik_1}\otimes\cdots\otimes
d_{ik_n}|(\nabla_{il_1},\cdots,\nabla_{il_n})>:=\delta_{k_1l_1}
\times\cdots{\times} \delta_{k_nl_n} \end{equation}
{\it and linear extension}.\vspace{0.5cm}

In a next step we extend these concepts to functions $f\in
C_0$ and {\it 'differential operators'} or {\it 'vector fields'} $\sum
a_{ik}\nik$. We have to check whether this is a natural(!) definition.\\[0.5cm]
{\bf 3.31 Observation}: {\it Vector fields $v:=\sum a_{ik}\nik$ are assumed
to act on functions $f=\sum f_in_i$ in the following manner}:
\begin{equation}v(f):=\sum a_{ik}(f_k-f_i)n_i\end{equation}
{\it i.e. they map $C_0\to C_0$}.\\[0.5cm]
{\bf 3.32 Corollary}: {\it Note that this implies}:
\begin{equation}\nik n_k=n_i\quad \nik n_i=-n_i\end{equation}
\begin{equation}\nki n_k=-n_k\quad \nki n_i=n_k\end{equation}\\[0.5cm]
{\bf 3.33 Observation}: {\it `Differential forms' $\omega =\sum
g_{ik}d_{ik}$ act on vector fields $v=\sum a_{ik}\nik$ according to}:
\begin{equation}<\omega|v>=\sum
  g_{ik}a_{ik}n_i\end{equation}\vspace{0.5cm}

With these definitions we can calculate $<df|v>$ with
\begin{equation}df=\sum
  (f_k-f_i)d_{ik}\end{equation}
Hence:
\begin{equation}<df|v>=<\sum (f_k-f_i)d_{ik}|\sum a_{ik}\nik>=\sum
  (f_k-f_i)a_{ik}n_i\end{equation}
which equals:
\begin{equation}(\sum a_{ik}\nik)(\sum
  f_in_i)=v(f)\end{equation}\\[0.5cm]
{\bf 3.34 Consequence}: {\it Our geometric interpretation of the algebraic
objects reproduces the relation}:
\begin{equation}<df|v>=v(f)\end{equation}
{\it known to hold in ordinary differential geometry, as is the case for
the following relations, and shows that the definitions made above
seem to be natural. Furthermore, vector and covector fields are left modules under the
action of ${\cal A}$}:
\begin{equation}(\sum f_in_i)(\sum a_{ik}\nik):=\sum f_ia_{ik}\nik
\end{equation}
\begin{equation}(\sum f_in_i)(\sum g_{ik}d_{ik}):=\sum
f_ig_{ik}d_{ik}\end{equation}\\[0.5cm]
{\bf 3.35 Conclusion}: {\it The above shows that, in contrast to classical
  differential geometry, we have a dual pairing between vector and
  covector fields with the vector fields acting as `twisted'
  derivations on the node functions while the corresponding
  differential forms obey the graded Leibniz rule like their classical
  counterparts.}\vspace{0.5cm}

Another important geometrical concept is the notion of `{\it
  connection}' or `{\it covariant derivative}'. Starting from the
  abstract concept of (linear) connection in the sense of Koszul it is
  relatively straightforward to extend this concept to the
  non-commutative situation, given a `{\it finite  projective module}'
  over some algebra ${\cal A}$ (instead of the sections of a vector
  bundle over some manifold ${\cal M}$, the role of $\A$ being played
  by the functions over ${\cal M}$; see e.g. \cite{Connes} or
  \cite{Coque}, as to various refinements and improvements cf. e.g. 
\cite{Dubois} and further references given there).

Without going into any details we want to briefly sketch how the
concept of connection can be immediately implemented in our particular
model without referring to the more abstract work. We regard (in a
first step) a connection as a (linear) map from the fields of tangent
vectors to the tensor product of tangent vectors and dual differential
forms as defined above (and having certain properties).\\[0.5cm]
{\bf 3.36 Definition/Observation(Connection)}: {\it A field of connections,
$\Gamma$, is defined at each node $n_i$ by a (linear) map}:
\beq \nik\to\gamma_{kl}^j(n_i)\cdot \nabla_{ij}\otimes d_i^l\eeq
{\it where the index $i$ plays rather the role of the ''coordinate'' $n_i$,
the index $l$ is raised in order to comply with the summation
convention. The $\gamma_{kl}^j$'s are called `connection
  coefficients'. The corresponding `covariant derivative'
$\nabla$ obeys the relations}:\\
i)\beq \nabla(v+w)=\nabla(v)+\nabla(w)\eeq
ii)\beq \nabla(f\cdot v)=v\otimes df+\nabla(v)\cdot f\eeq
iii)\beq \nabla(\nik)=\Gamma(\nik)\quad
df=\sum(f_k-f_i)d_{ik}\eeq\\[0.5cm]
{\it Remark}: The tensor product in ii) is understood as the pointwise
product of fields at each node $n_i$, i.e. $\nabla_{ik}$ going with
$d_{ik}$. This is to be contrasted with the abstract notion of tensor
product in e.g. the above differential algebra $\Omega(\A)$ which does
{\bf not} act locally, the space consisting of, say, elements of the kind
$n_1\otimes n_2\otimes\cdots\otimes n_k$. These different parallel
structures over the same model shall be scrutinized in more detail
elsewhere. Note that the above extra locality structure is a
particular property of our model class and does not (openly) exist in
the general approach employing arbitrary {\it projective modules}
respectively {\it differential algebras}.

\section{Intrinsic Dimension in Networks, Graphs and other Discrete Systems}
There exist a variety of concepts in modern mathematics which extend
the ordinary or naive notion of {\it dimension} one is accustomed to
in e.g. differential topology or linear algebra. In fact, {\it
  topological dimension} and related concepts are notions which are
perhaps even closer to the underlying intuition (cf. e.g.
\cite{Edgar}).

Apart from the purely geometric concept there is also an important  physical
role being played by something like dimension, having e.g. pronounced effects
on the behavior of, say, many-body-systems near their {\it phase
transition points} or in the {\it critical region}.

But even in the case of e.g. lattice systems they are usually
treated as being embedded in an underlying continuous background space
(typically euclidean) which supplies the concept of ordinary
dimension so that the {\it intrinsic dimension} of the discrete array itself
does usually not openly enter the considerations.

Anyway, it is worthwhile even in this relatively transparent
situation to have a closer look on the situations where attributes of something
like dimension really enter the physical stage. Properties of
models of, say, statistical mechanics  are typically derived from
the structure of the microscopic interactions of their constituents.
This then is more or less the only place where dimensional aspects enter
the calculations.

Naive reasoning might suggest that it is something like the number of nearest
neighbors (in e.g. lattice systems) which encodes the dimension of the underlying space and influences via that way the
dynamics of the system. However, this surmise, as we will show in the
following, does not reflect the crucial point which is considerably
more subtle.

This holds the more so for systems which cannot be considered as being
embedded in a smooth regular background and hence do not inherit their
dimension from the embedding space. A case in point is our primordial
network in which Planck-scale-physics is assumed to take
place. In our approach it is in fact exactly the other way round:
Smooth space-time is assumed to emerge via a {\it phase transition} or a
certain {\it cooperative behavior} and
after some `{\it coarse graining}' from this more fundamental
structure.\\[0.5cm]
{\bf 4.1 Problem}: {\it Formulate an intrinsic notion of dimension for
model theories without making recourse to the dimension of some
continuous embedding space}.\vspace{0.5cm}

In a first step we will show that graphs and networks as introduced
in the preceding sections have a natural metric structure. We have
already introduced a certain neighborhood structure in a graph with
the help of the minimal number of consecutive bonds connecting two
given nodes.

In a connected graph any two nodes can be connected by a sequence of
bonds. Without loss of generality one can restrict oneself to {\it
paths}. One can then define the length of a path (or sequence of
bonds) by the number $l$ of consecutive bonds making up the
path.\\[0.5cm] 
{\bf 4.2 Observation/Definition}: {\it Among the paths connecting two
arbitrary nodes there exists at least one with minimal length which
we denote by $d(n_i,n_k)$. This d has the properties of a metric, i.e:}
\begin{eqnarray} d(n_i,n_i) & = & 0\\ d(n_i,n_k) & = &
d(n_k,n_i)  \\d(n_i,n_l) & \leq & d(n_i,n_k)+d(n_k,n_l) \end{eqnarray}
(The proof is more or less evident).\\[0.5cm]
{\bf 4.3 Corollary}: {\it With the help of the metric one gets a natural
neighborhood structure around any given node, where ${\cal U}_m(n_i)$
comprises all the nodes, $n_k$, with $d(n_i,n_k)\leq m$, 
$\partial{\cal U}_m(n_i)$
the nodes with $d(n_i,n_k)=m$}. \vspace{0.5cm}

This natural neighborhood structure enables us to develop
the concept of an intrinsic dimension on graphs and networks. To this
end one has at first to realize what property really matters
physically (e.g. dynamically), independently of the kind of model or embedding
space. \\[0.5cm]
{\bf 4.4 Observation}: {\it The crucial and characteristic property of,
say, a graph or network which may be associated with something like
dimension is the number of `new nodes' in ${\cal U}_{m+1}$ compared
to ${\cal U}_m$ for m sufficiently large or $m\to \infty$. The deeper
meaning of this quantity is that it measures the kind of `wiring' or `connectivity' in the network and is therefore a
`topological invariant'}.\\[0.5cm]
{\it Remark}: In the following I shall be very brief on this
ramified topic as much more details have been presented and
proved in a separate paper (\cite{Req1}) to which the interested
reader is referred. Instead of that I would like to discuss certain
aspects and ideas being related to this concept.\vspace{0.5cm}

In many cases one expects the number of nodes in ${\cal U}_m$ to grow like
some power $D$ of $m$ for increasing $m$. By the same token one expects the
number of new nodes after an additional step to increase proportional
to $m^{D-1}$. With $|\,\cdot\,|$ denoting the number of nodes
we hence expect frequently the following to hold:
\begin{equation} |{\cal U}_{m+1}|-|{\cal U}_m|=|\partial {\cal U}_{m+1}|=f(m)
\end{equation}
with
\begin{equation} f(m)\sim m^{D-1} \end{equation}
for $m$ large.\\[0.5cm]
{\bf 4.5 Definition}: {\it ``The'' intrinsic dimension D of a homogeneous 
(infinite) graph is given by} 
\begin{equation} D-1:=\lim_{m\to \infty}(\ln f(m)/\ln m)
\end{equation}
{\it or}
\begin{equation} D:=\lim_{m\to \infty}(\ln |{\cal U}_m|/\ln m)
\end{equation}
{\it provided that a unique limit exists!\\
What does exist in any case is $\liminf$ respectively $\limsup$ which
can then be considered as upper and lower dimension. If they coincide
we are in the former situation. By `homogeneous' we mean that 
the graph ``looks locally more or less the same'' geometrically everywhere, in
particular its dimension $D$ should not depend on the reference point
(and that the two definitions of graph dimension given above
coincide!).}
\\[0.5cm]
{\it Remarks}:
\begin{enumerate}
\item One might naively expect that `{\it regularity}', i.e. constant
node degree, plus certain other conditions imply homogeneity but this is
a highly non-trivial question. There are e.g. simple examples of
regular graphs which do not ``look the same'' around every node. That
is, a so-called `{\it trivalent graph}', which is frequently taken as
a discretized standard example in various versions of quantum gravity
(see e.g. the above mentioned literature about spin networks), is
`{\it planar}', i.e. can be embedded in the plane. This may lead to the
erroneous conclusion that its ``natural dimension'' is two. On the
other side, our analysis shows, that its ``intrinsic dimension''
depends crucially on the details of its wiring and can even become
infinite in case of a trivalent tree. Furthermore, it is a fascinating
and non-trivial task to characterize e.g. the dimension
of regular lattices (or `{\it triangulations}') in a purely
combinatorial manner
without using a background space. This shall however be done elsewhere.
\item  On the other side we showed in \cite{Req1} that, among other things, uniform 
boundedness of the node 
degree guarantees already the independence of $D$ with respect to the
reference point! Furthermore we were able to construct graph models
for each given (fractal) dimension $D\in\Rl$. Due
to discreteness, the relation between the two definitions of
dimension may be a little bit subtle (as is the case for the various
fractal dimensions; there may be exceptional graphs where they do not 
coincide!). 
\item Other (but related) definitions
of dimension are possible,  incorporating e.g. the bonds instead of the nodes.
\item For practical purposes one may also introduce a notion of local
dimension around certain nodes or within certain regions of a regular graph if the above limit is approached
sufficiently fast locally.
\item Particularly interesting phenomena are expected to show up if
  this concept is applied within the regime of `{\it random graphs}'
  (see e.g. \cite{Req4})
\item For some remarks concerning more or less related ideas being
  scattered in the literature (known to us) see \cite{Req1}
\end{enumerate}
\vspace{0.5cm}
  
That this definition is reasonable can be seen by applying it to
ordinary cases like regular translation invariant lattices. It is
however not evident that such a definition makes sense for arbitrary
graphs, in other words, a (unique) limit point may not always
exist. It seems to be a highly non-trivial task to characterize
the conditions which imply such a limit to exists. 
\\[0.5cm]
{\bf 4.6 Observation} {\it For regular lattices D coincides with the
dimension of the euclidean embedding space $D_E$}.\\[0.5cm]
Proof: This can be rigorously proved but it is perhaps more
instructive to simply draw a picture of the consecutive series
of neighborhoods of a fixed node for e.g. a 2-dimensional Bravais
lattice. It is fairly obvious that for m sufficiently
large the number of nodes in $\U_m$ goes like a power of m with the
exponent being the embedding dimension $D_E$ as the euclidean volume
of $\U_m$ grows with the same power.\\[0.5cm]
{\it Remarks}:
\begin{enumerate}
\item For $\U_m$ too small the number of nodes may deviate from
an exact power law which in general becomes only correct for
sufficiently large $m$.
\item The number of nearest neighbors, on the other side, does {\it not}
influence the exponent but rather shows up in the prefactor. In other
words, it influences $|\U_m|$ for $m$ small but drops out
asymptotically by taking the logarithm. For an ordinary Bravais
lattice with $N_C$ the number of nodes in a unit cell one has
asymptotically:
\begin{equation} |\U_m|\sim N_C\cdot m^{D_E} \quad\mbox{and hence:}
\end{equation}
\begin{equation} D=\lim_{m\to\infty}(\ln(N_C\cdot m^{D_E})/\ln
m)=D_E+\lim_{m\to\infty}(N_C/\ln m)=D_E 
\end{equation}
independently of $N_C$.\end{enumerate}  \vspace{0.5cm}

Matters become much more interesting and subtle if one studies more
general graphs than simple lattices. Note that there exists a general
theorem (in fact a specialisation of the more general result about
simplicial complexes we mentioned in the preceeding section to one-simplices), showing that practically every graph can be embedded in
$\Rl^3$ and still quite a few in $\Rl^2$ ({\it planar graphs}). 

This shows again that something like the dimension of the embedding
space is in general not a characteristic property of a network or
graph. Quite to the contrary, it is generically entirely uncorrelated
with its {\it intrinsic dimension} we defined above.
An extreme example is a `{\it tree graph}', i.e. a graph without
`{\it loops}'. In the following we study an infinite, regular tree
graph with node degree 3, i.e. 3 bonds branching off each node. The
absence of loops means that the `{\it connectivity}' is extremely low
which results in an exceptionally high `{\it dimension}' as we will
see.

Starting from an arbitrary node we can construct the neighborhoods
$\U_m$ and count the number of nodes in $\U_m$ or $\partial\U_m$.
$\U_1$ contains 3 nodes which are linked with the reference node
$n_0$. There are 2 other bonds branching off each of these nodes.
Hence in $\partial\U_2=\U_2\backslash\U_1$ we have $3\cdot2$ nodes
and by induction:
\begin{equation} |\partial\U_{m+1}|=3\cdot2^m \end{equation}
which implies
\begin{equation} D-1:=\lim_{m\to\infty}(\ln|\partial\U_{m+1}|/\ln m)=
\lim_{m\to\infty}(m\cdot ln 2/\ln m+3/\ln m)=\infty \end{equation}
Hence we have: \\[0.5cm]
{\bf 4.7 Observation(Regular Tree)}: The intrinsic dimension of an infinite
regular tree is $\infty$ and the number of new nodes grows exponentially like
some $n(n-1)^m$  (with $n$ being the node degree).\\[0.5cm]
{\it Remark}: $D=\infty$ is mainly a result of the absence of loops and of regularity, in
other words: there is exactly one path, connecting any two nodes. 
This is usually not so in other graphs, e.g. lattices, where the
number of new nodes grows at a much slower pace (whereas the number
of nearest neighbors can nevertheless be large). This is due to the
existence of many loops s.t. many of the nodes which can be reached
from, say, a node of $\partial\U_m$ by one step are already
contained in $\U_m$ itself. On the other side we recently constructed
(\cite{Req1}) tree graphs having an arbitrary
fractal dimension (they are however not regular, i.e. having a
varying node degree, in particular they have a lot of end
points). Hence the absence of loops is not the only reason for large
graph dimensions.\vspace{0.5cm}

We have seen that for, say, lattices the number of new nodes grows
like some fixed power of $m$ while for, say, regular trees $m$ occurs in the
exponent. The borderline can be found as follows:\\[0.5cm]
{\bf 4.8 Observation}: If for $m\to\infty$ the average number of 
nodes in $\U_{m+1}$ per node contained in $\U_m$ is uniformly away
from zero or, stated differently:
\begin{equation} |\U_{m+1}|/|\U_m|\geq 1+\varepsilon \end{equation}
for some $\varepsilon\geq 0$ then we have exponential growth, in other
words, the intrinsic dimension is $\infty$.
\\The corresponding result holds with $\U_m$ being replaced by
$\partial\U_m$. \\[0.5cm]
Proof: If the above estimate holds for all $m\geq m_0$ we have by
induction:
\begin{equation} |\U_m|\geq |\U_{m_0}|\cdot (1+\varepsilon)^{m-m_0}
\end{equation} 
\vspace{0.5cm}

Potential applications of this concept of intrinsic dimension are
manifold. Our main goal is it to develop a theory which explains how
our classical space-time and what we like to call the '{\it physical vacuum}'
has emerged from a more primordial and discrete background via some
sort of phase transition (the first preliminary steps being done in \cite{Req4}).

In this context we can also ask in what sense {\it macroscopic} space-time dimension 4
is exceptional, i.e. whether it is merely an accident or whether
there is a cogent reason for it. We hope or expect that the above
concept of intrinsic dimension together with the dynamics of its
possible change may help to come to grips with such a fundamental question.

\end{document}